# THE HABITABLE ZONES OF PRE-MAIN-SEQUENCE STARS

Ramses M. Ramirez[1,2,3] and Lisa Kaltenegger[1,2]

## ABSTRACT

We calculate the pre-main-sequence HZ for stars of spectral classes F–M. The spatial distribution of liquid water and its change during the pre-main-sequence phase of protoplanetary systems is important in understanding how planets become habitable. Such worlds are interesting targets for future missions because the coolest stars could provide habitable conditions for up to 2.5 billion years post-accretion. Moreover, for a given star type, planetary systems are more easily resolved because of higher pre-main-sequence stellar luminosities, resulting in larger planet-star separation for cool stars than is the case for the traditional main-sequence (MS) habitable zone (HZ). We use 1-D radiative-convective climate and stellar evolutionary models to calculate pre-main-sequence HZ distances for F1-M8 stellar types. We also show that accreting planets that are later located in the traditional MS HZ orbiting stars cooler than a K5 (including the full range of M-stars) receive stellar fluxes that exceed the runaway greenhouse threshold, and thus may lose substantial amounts of water initially delivered to them. We predict that M-star planets need to initially accrete more water than Earth did or, alternatively, have additional water delivered later during the long pre-MS phase to remain habitable. Our findings are also consistent with recent claims that Venus lost its water during accretion.

[1]Institute for Pale Blue Dots, Cornell University, Ithaca, NY, USA
[2]Department of Astronomy, Cornell University, Ithaca, NY, USA
[3]Center for Radiophysics and Space Research, Cornell University, Ithaca, NY, USA

## 1. INTRODUCTION

Ground- and space-based searches for extrasolar planets have found 1822 planets and more than 2900 further candidates (October 2014)(e.g. Batalha et al.2013, Mayor and Queloz, 2012). One of the aims of the search, and a dedicated goal for NASA's *Kepler spacecraft,* is to determine the frequency of rocky planets within the habitable zone (HZ) (e.g. Borucki et al.2011), the circumstellar region where water could be liquid on the surface of rocky planets (e.g., Kasting et al.1993). Planets in the HZ of cool stars are easier to detect for current as well as upcoming searches like NASA's all sky Transiting Exoplanet Survey Satellite (TESS) (Ricker et al.2014) because of increased transit probability, transit frequency, and higher signal-to-noise for both transit and radial velocity detections at a given planetary size and mass. Also, such planets provide better targets for spectroscopic characterization by the next-generation Thirty-Meter Telescope (TMT), Giant Magellan Telescope (GMT), and European Extremely Large Telescope (E-ELT) for nearby stars (see e.g. Kaltenegger & Traub 2010). In addition, the calculated fraction of planets is high (between 40 and 60%)(Bonfils et al.2013, Dressing et al.2013, Kopparapu et al.2013, Gaidos 2013).

Previous studies have focused on the MS HZ (i.e. Kasting et al.1993; Kopparapu et al.2013,2014), the effect of metallicity on the HZ (Danchi & Lopez 2013), and (recently) potential abiotic $O_2$ build-up in the atmosphere (Luger & Barnes 2014 submitted). However the pre-MS period of M-stars can last up to 2.5 billion years, potentially providing habitable conditions for nearly that long. In addition, understanding the pre-MS stage is crucial to water loss and delivery for both pre-MS HZ and MS HZ planets. Here, we calculate the pre-MS HZ of F to M-stars to assess if this region provides additional targets in the search for habitable planets. The pre-MS HZ is wider than their MS HZ so young planets orbiting close-by cool stars could be resolved with future ground- and space-based telescopes.

The question of a planet's habitability is linked to its ability to acquire and retain water. Former arguments against M-star planet habitability focused on the difficulty that planetesimals in M-star HZs would have in acquiring volatiles. This was either attributed to higher orbital speeds, leading to more energetic collisions with other bodies (Lissauer 2007), or they began dry because of inefficient radial mixing, and so fewer volatile-rich planetesimals from larger distances would be accreted (Raymond et al.2007). These concerns have been recently questioned by the in-situ accretion model of Hansen (2014), which predicts gatekeeper planetesimals may be large enough to reduce collisional velocities, allowing M-star planets to acquire many hundreds of times Earth's surface water endowment.

A problem, which has been neglected, concerns volatile *retention* (Fig.2). Pre-MS luminosities for M-stars are much higher relative to their zero-age-main-sequence (ZAMS) values than for F or G-stars. An M8-star can be about 180 times as bright as its ZAMS luminosity during its contraction stage, whereas the Sun was about twice as bright (Baraffe et al.1998; 2002, Fig.1). High stellar fluxes during planetary accretion can prove detrimental to volatile retention, particularly when insolation levels exceed the runaway greenhouse threshold (see Fig.2 Kasting, 1988; Abe, 1988; Goldblatt et al.2013; Ramirez et al.2014).

We describe our models in section 2, calculate the pre-MS HZ and discuss water delivery in section 3. Results are discussed in section 4, followed by concluding comments.

## 2. METHODS
*2.1 The pre-main-sequence stellar grid*

We compute the pre-MS luminosities of a grid of representative stars (Fig.1) using stellar evolution models (Baraffe et al.1998, 2002). The corresponding masses for spectral classes F1 to M8 are: $1.5M_{sun}$, $1.3M_{sun}$, $1M_{sun}$, $0.75M_{sun}$, $0.5M_{sun}$, $0.15M_{sun}$, and $0.08M_{sun}$. These spectral classes are consistent with the star's age at 4.5 Gyr (2 Gyr for the F1). All calculations commence after a proto-planetary disk age of 1 million years because uncertain initial conditions render earlier computations unreliable (Baraffe et al.2002). For stars cooler than about 3000K, the initial luminosity is very high and decreases over time until the star reaches the ZAMS. Hotter stars show a brightening just before the ZAMS.

Approximate accretion time scales for HZ planets are estimated from models by Raymond et al.(2007). We use accretion times of 50, 40, 20, 8, and 5 million years for the Sun, K5, M1, M5 and M8-stars respectively. These timescales span the entire accretion period from planetesimal growth to final assembly (e.g. Chambers 2004). The accretion time for the M8 star was obtained by extrapolation.

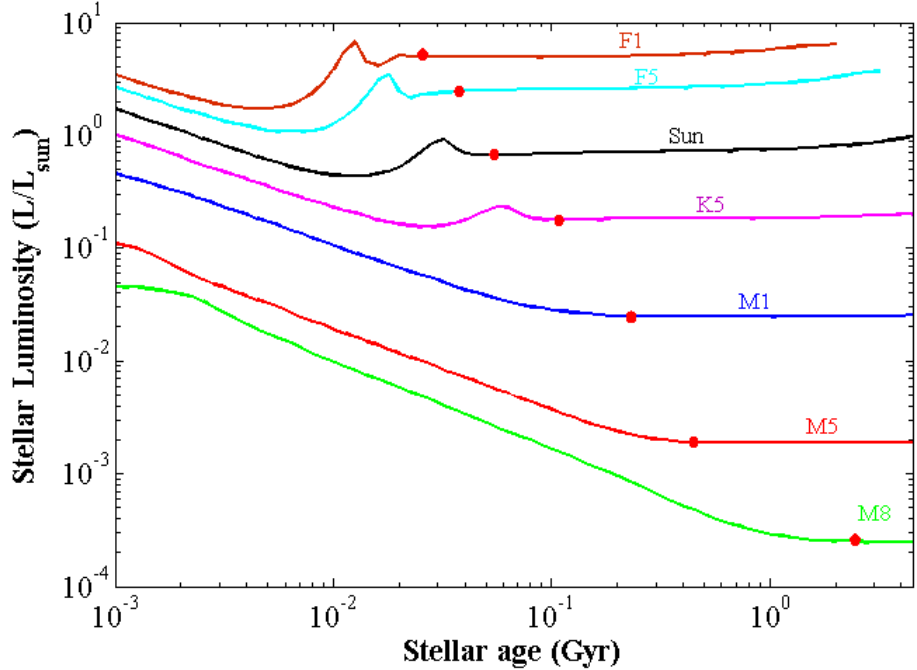

**Fig.1:** Evolution of stellar luminosity for our grid of F-M stars (F1, F5, Sun, K5, M1, M5, and M8)(Baraffe et al.1998,2001). When the star reaches the main-sequence (red point) its luminosity curve flattens.

*2.2 HZ models*

We use the effective stellar flux, $S_{eff}$, at the empirical MS HZ boundaries (Kasting 1993, Kopparapu et al.2014) to calculate the pre-MS inner and outer HZ limits. $S_{eff}$, is defined as the stellar flux divided by the solar flux at Earth's orbit. The inner edge is defined by the stellar flux received by Venus when we can exclude the possibility that it had standing water on the surface (about 1Gyr ago), equivalent to a stellar flux of $S_{eff}$=1.77 (Kasting et al.1993). The outer edge is defined by the stellar flux that Mars received at the time that it may have had stable water on its surface (about 3.8Gyr ago), which corresponds to $S_{eff}$=0.32. In addition, Fig.2 to 4 also show a more conservative inner edge of the HZ (dashed lines) based on 3D models ($S_{eff}$~1.11) (Leconte et al.2013). However, model properties, including clouds, are still evolving and their applicability to water-rich atmospheres (i.e. inner edge of the HZ) are not fully understood yet. Therefore, we use the empirical HZ limits for our baseline calculations and discuss how our results would change using the inner edge model limit (section 4.1).

Further potential inward extension of the inner edge caused by tidal locking (Yang et al.2013;2014) is neglected because such young planets would not have had time to spin down. The initial luminosities are ~ 2/3, 2.4, 4, and 180 times their ZAMS values for the F1, Sun, K5 and M8-stars, respectively (Baraffe et al.2002). The changing stellar effective temperature and stellar energy distribution (SED) throughout the pre-MS impacts the relative contribution of absorption versus scattering of the protoplanetary atmosphere, which changes $S_{eff}$ at the HZ boundaries through time (see Fig.2).

Here we focus on Earth-sized planets and start with an initial water inventory equivalent or higher to that of Earth for a 1-bar Nitrogen atmosphere that is water-dominated at the inner edge and $CO_2$-dominated near the outer edge of the HZ (following Kasting et al.1993). Thus, we do not consider HZ limits for extremely dry (Abe et al.2011, Zsom et al.2014), or hydrogen-rich atmospheres (Pierrehumbert and Gaidos, 2011; see Kasting et al.2013 for critical discussion of both limits) in this paper. For Earth-like planets, the runaway greenhouse state (complete ocean evaporation) would be triggered when the planet's surface temperature reaches the critical temperature for water (647K) (Kasting et al.1993; Ramirez et al.2014).

## 3. RESULTS
### 3.1 Pre-Main-Sequence Habitable Zone limits

The orbital distance of the HZ changes throughout the pre-MS stage of the host star. Both distance limits evolve due to the star's changing luminosity and SED. The inner edge of the pre-MS HZ (see Fig.2) for an F1-star is initially located at 2AU at the beginning and moves in to 1.4AU by the end of its pre-MS stage (about 25 million years). The outer edge evolves from 4.2AU to 2.5AU during the same time. For the coolest grid star (M8), the inner edge is initially located at 0.16AU at the beginning and moves to 0.01AU at the end of the pre-MS stage after about 2.5 billion years. The outer edge of the HZ moves from 0.45AU to 0.3AU through the same time. For the Sun, the inner edge of the HZ changes from 1AU at the beginning to 0.6AU at the end of the pre-MS stage after about 50 million years. The outer edge correspondingly moves from 2.6AU to 1.5AU. Approximate pre-MS HZ boundaries for stars with lifetimes >200Myr (0.08–0.5$M_{sun}$) are given in eqn. 1.

$$d(AU) = at^b + c \qquad (1)$$

where, t is time since stellar birth (in Myr) and a,b,and c are constants given in table1.

Although how planetary mass scales with HZ remains poorly understood, mass variations have only a small effect according to a recent model (Kopparapu et al.2014). For both super-Earths (5 Earth masses) and Mini-Earths (0.5 Earth masses) the inner HZ orbital distance decreases (with increasing) or increases (with decreasing planetary mass) by a maximum of 4%, regardless of spectral class.

To explore the effect of the high pre-MS stellar luminosity on water-bearing planetesimals, we compute a pre-MS ice-line distance scaled from the present day Solar System location using stellar luminosity ($\sim(T_{star}/5800 \text{ K})^2 * 2.5 * (M_{star}/M_{sun})$) (following Hansen et al.2014). We use the corresponding $S_{eff}$ to scale the pre-MS ice-line distance backwards in time ($\sim(L/S_{eff})^{1/2}$).

The HZ boundaries and iceline both change substantially throughout the star's pre-MS stage (Fig. 2). This is most pronounced for cool stars because of their high initial luminosity. A planet needs to orbit its host star farther out than the Runaway Greenhouse limit (i.e. inner edge of the HZ) to retain its water throughout the star's evolution (see section 3.3) whereas a planetesimal needs to reside beyond the iceline to do the same. Particularly for cold stars, planetesimals have to orbit much farther away to retain their water initially than one would expect based on a star's MS luminosity.

### 3.2 Stellar flux at Main-Sequence HZ exceed runaway greenhouse limits during pre-MS

The stellar fluxes ($S_{eff}$) received at the boundaries of the MS HZ (Kasting et al.1993, Kopparapu et al.2013,2014) and the ice line during the star's pre-MS phase are shown in Fig.3. Regardless of star type, all planets located at the inner edge of the MS HZ receive stellar fluxes that exceed the runaway greenhouse threshold for at least part of the pre-MS stage (Fig.3), which results in rapid evaporation of surface water.

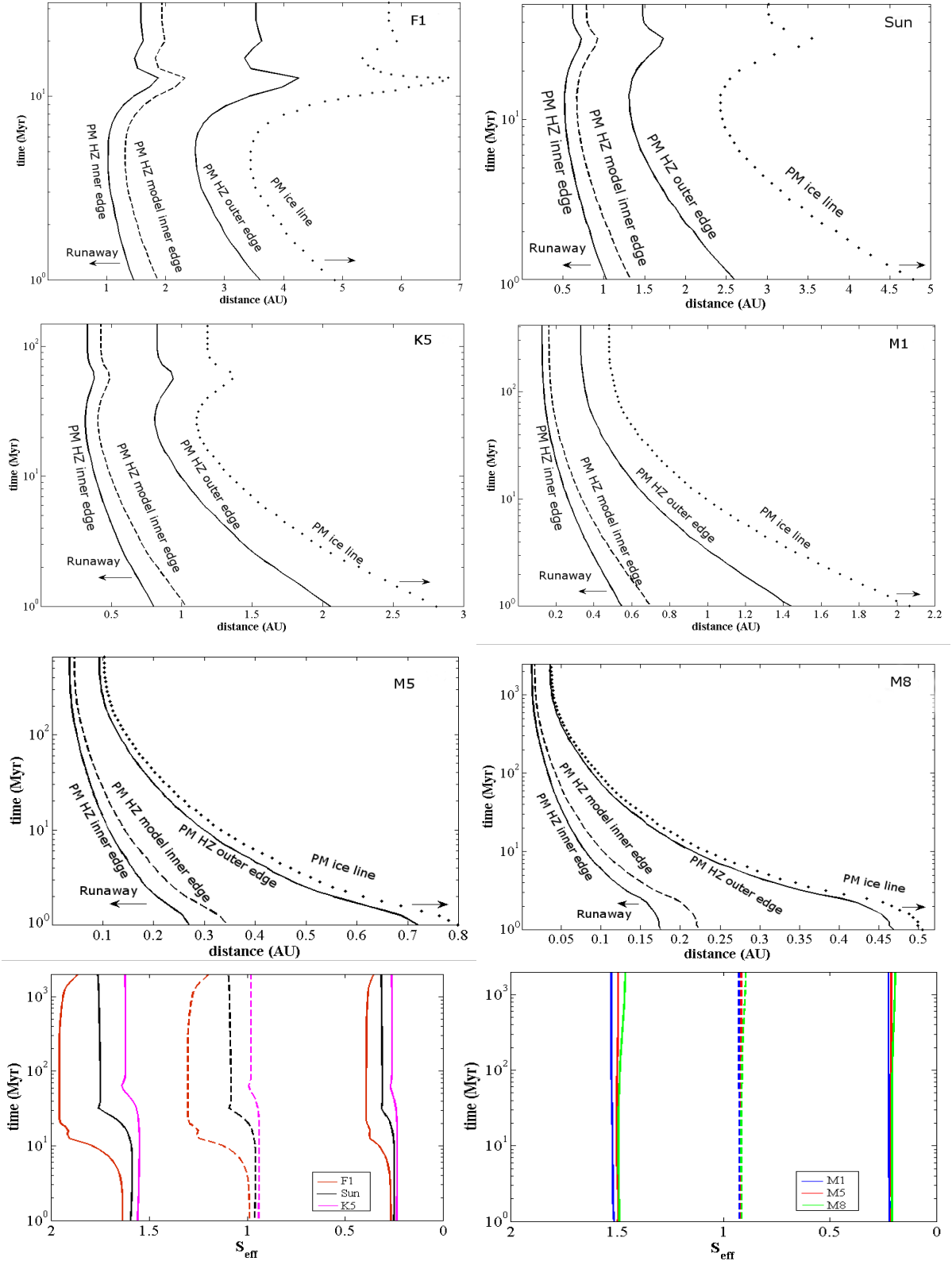

**Fig.2:** Pre-MS (PM) empirical HZ boundaries (solid), modeled boundaries (dashed), and ice lines (dotted curves) for pre-MS stars (F1 to M8) in distance and effective stellar flux. Planets become too hot and get devolatilized for distances inside the inner edge of the HZ (i.e. Runaway Greenhouse threshold).

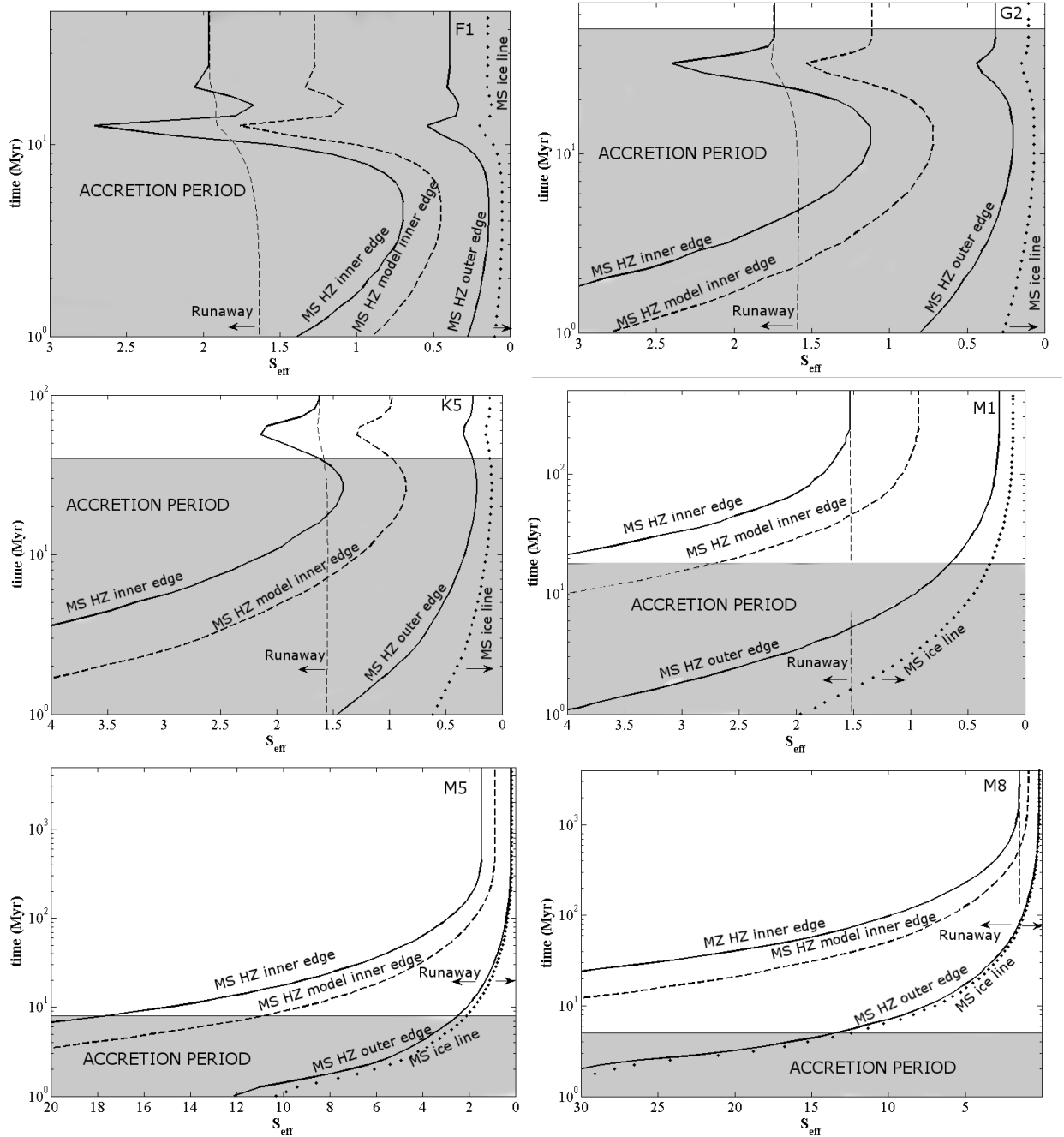

**Fig.3**. Stellar flux ($S_{eff}$) received during the stars pre-main-sequence phase at the empirical MS HZ limits (solid curves), model inner edge limit (dashed curves), and ice line (dotted curve) (F1 to M8-stars). Planets become too hot and get devolatilized for $S_{eff}$ values that exceed the runaway greenhouse threshold (vertical dashed line). Approximate accretion timescales are shown as shaded regions (following Raymond et al.2007).

The pre-MS luminosities of the coolest grid star (M8) to its MS value are about 75 times higher than is the corresponding luminosity change for the Sun, with incident stellar fluxes exceeding 20 times the value at Earth's orbit and 10 times that received by Venus for M8 host stars at the inner edge of the MS HZ (Fig.4). The M-star fluxes during the pre-MS HZ are high enough to trigger runaway greenhouse conditions beyond the MS HZ outer edge, and even into the water-rich planetesimals located beyond the ice line (Fig.3). By contrast, only F-star planets orbiting near the MS HZ inner edge receive stellar fluxes that exceed the runaway greenhouse threshold for any substantial amount of time.

*3.3 Loss of oceans and hydrogen to space*

Triggering a runaway greenhouse does not ensure that a planet could not later become habitable because water could recondense afterward. Whether water is permanently lost depends on how much hydrogen escapes from the top of the atmosphere. Here, we assume that photolysis reactions in the atmosphere would have efficiently converted $H_2O$ to H (following Kasting & Pollack 1983) and that the escape of H is energy-limited (Watson et al.1981). This assumption is reasonable because in a runaway greenhouse atmosphere, if water is the dominant volatile, the $H_2O$ mixing ratio at the cold trap should be large enough to overcome the diffusion limit (Walker, 1977). Under energy-limited conditions, the escape rate (in moles/second), $F_{el}$, is given by eqn. 2 (Abe et al.2011):

$$F_{el} \approx \frac{4\pi R^3 \varepsilon S_{EUV}}{GMm_h} \quad (2)$$

where $\varepsilon$ (=0.3) is a nominal value for the heating efficiency (following Watson et al.1981, Lammer et al.2009), $S_{EUV}$ the stellar extreme ultraviolet (EUV) flux, G the gravitational constant, M and R the planetary mass and radius (here set to Earth's values), and $m_h$ the mass of an H atom. The globally-averaged $S_{EUV}$ flux incident on Earth today is $\simeq 1.16$ ergs/cm$^2$/sec, which is over 100 times lower than its value >4.5Gya (Penz et al.2008). By integrating eqn.1 and using EUV parameterizations (Lammer et al.2009), we can estimate the water losses for planets around F-M stars. Note that calculated losses for planets orbiting M8 stars may be somewhat overestimated because $S_{EUV}$ fluxes are high enough (>~10,000 ergs/cm$^2$/sec) for Lyman-alpha cooling to become important, which is neglected in eqn.2 (Murray-Clay et al. 2009).

$$Moles(H_2)lost = \int F_{el}\, dt = B\frac{4\pi R^3 \varepsilon}{9GMm_h}\int_{t_i}^{t_f} S_{EUV}\, dt;$$

$$S_{EUV} = \frac{L}{4\pi d^2};$$

$F-$ stars:
$L = 0.284 L_o t^{-0.547}, L_o = 10^{29.83} ergs/\sec$

$G-$ stars:
$L = 0.375 L_o t^{-0.425}, L_o = 10^{29.35} ergs/\sec \quad (3)$

$K-$ stars:
$L = 0.474 L_o t^{-0.324}, L_o = 10^{28.87} ergs/\sec$

$M-$ stars:
$L = 0.17 L_o t^{-0.77}, t \leq 0.6 Gyr$
$L = 0.13 L_o t^{-1.34}, t > 0.6 Gyr, L_o = 10^{28.75} ergs/\sec$

where L is the time-varying stellar luminosity (ergs/sec), $L_o$ the stellar luminosity at 0.1Ga, d the orbital distance, and B the number of seconds per Gyr. Although stellar luminosity measurements before about 0.1Ga are unavailable (e.g. Ribas et al.2005; Lammer et al.2009), we extrapolate these relationships to 1 Myr. Thus, this assumes that EUV fluxes continue to increase with luminosity. For a given planet, eqn. 3 yields the total number of moles of $H_2$ that escape to space from the start of accretion ($t_i$ =1 million years) to the stellar age beyond which incident fluxes decrease below the runaway greenhouse threshold ($t_f$). A threshold crossover mass (e.g. eqn.16 of Hunten & Pepin 1987) of 16g/mol (the mass of O) is always exceeded during runaway conditions for rocky planetary masses for all grid stars. This results in escape of O as well, which makes H losses only $(2/(2+16))$ $1/9$ as efficient (eq. 2). Thus, escape in this regime drags both H and O away, reducing the loss of H proportionately (Lammer et al, 2009). The mass of $H_2$ in 1 Earth ocean is $((2/18)*1.4x10^{24}g)$ $1.56x10^{23}g$ or $7.8x10^{22}$ moles of $H_2$. The number of oceans lost is then obtained by dividing the moles of $H_2$ lost by $7.8x10^{22}$.

Initial stellar incident fluxes and potential water loss in the MS HZ region during the pre-MS phase of the parent star are greatest for cool stars. Planets orbiting at distances corresponding to the inner and outer MS HZ around an M8-star would lose up to 3800 and 225 oceans, respectively, 350 and 15 oceans around an M5, and 25 and 0.5 oceans around an M1. Planets orbiting at the outer edge of the MS HZ for stars from K5 and hotter would retain their water. Planets orbiting at distances corresponding to the inner edge of the MS HZ for K5 and G2-stars would lose about 1.3 and 0.75 Earth oceans, respectively. In contrast, water loss for planets in the MS HZ around F-stars is negligible during the pre-MS phase of the host star. At the MS iceline, about 200, 12, and 0.1 oceans are lost for M8, M5, and M1 host stars, respectively, suggesting that even migrating water-rich planetesimals in mid- to late-M-star systems are susceptible to volatile loss. However, these losses decrease with increasing orbital distance. To estimate losses for different planetary mass, crossover mass and water losses are appropriately scaled by the mass, radius, and gravity.

*3.4 Stellar flux received at the location of Earth, Venus and Mars during the pre-main-sequence*

We assess the maximum water loss for rocky planets orbiting our grid host stars during the pre-MS stage at the Venus-equivalent, Earth-equivalent, and Mars-equivalent orbital distances (scaled to the stellar fluxes at 0.72, 1, and 1.52AU in our Solar System)(Fig.4). For the Sun, Mars and Earth never receive stellar fluxes high enough to trigger runaway conditions in our model. This is consistent with suggestions that Earth obtained most of its water inventory during accretion (see e.g. Elkins-Tanton 2011; Gomes et al.2005). For the hottest grid star (F1) both Earth- and Mars-analogs are stable against runaway while Venus-analogs experience runaway conditions for a few Myr. In contrast, around the coolest grid star (M8) Earth- and Mars-analogs could lose up to 2300 and 800 oceans, respectively, with Venus-analogs remaining in an indefinite runaway state. We calculate that at least one Earth ocean is lost for Venus-, Earth-, and Mars-analogs around stars cooler than a G8, K8, and M4, respectively.

## 4. DISCUSSION
*4.1 Results for modeled MS HZ limit*

The dashed lines in Fig.2 to Fig.4 show the stellar flux, orbital distance and runaway threshold for the modeled HZ inner edge. Planets orbiting at distances corresponding to these limits are farther away from the star, resulting in up to 2400, 220, 8.5, 0.75, and 0.7 Earth oceans lost during the pre-MS phase for the M8, M5, M1, K5, and G2-stars, respectively (Fig.4). In contrast, the modeled inner edge limit encroaches closer to the HZ planets, which increases their runaway instability. We calculate that at least one Earth ocean is lost for Venus-, Earth-, and Mars-analogs around stars cooler than a G4, K5, and M0 respectively. The results for our Sun suggest that Venus would have experienced runaway conditions during most of its accretionary period, as posited by Hamano et al.(2013). In this scenario, the leftover oxygen would have been consumed by the surface magma ocean, which could explain the lack of oxygen measured in the current Venusian atmosphere.

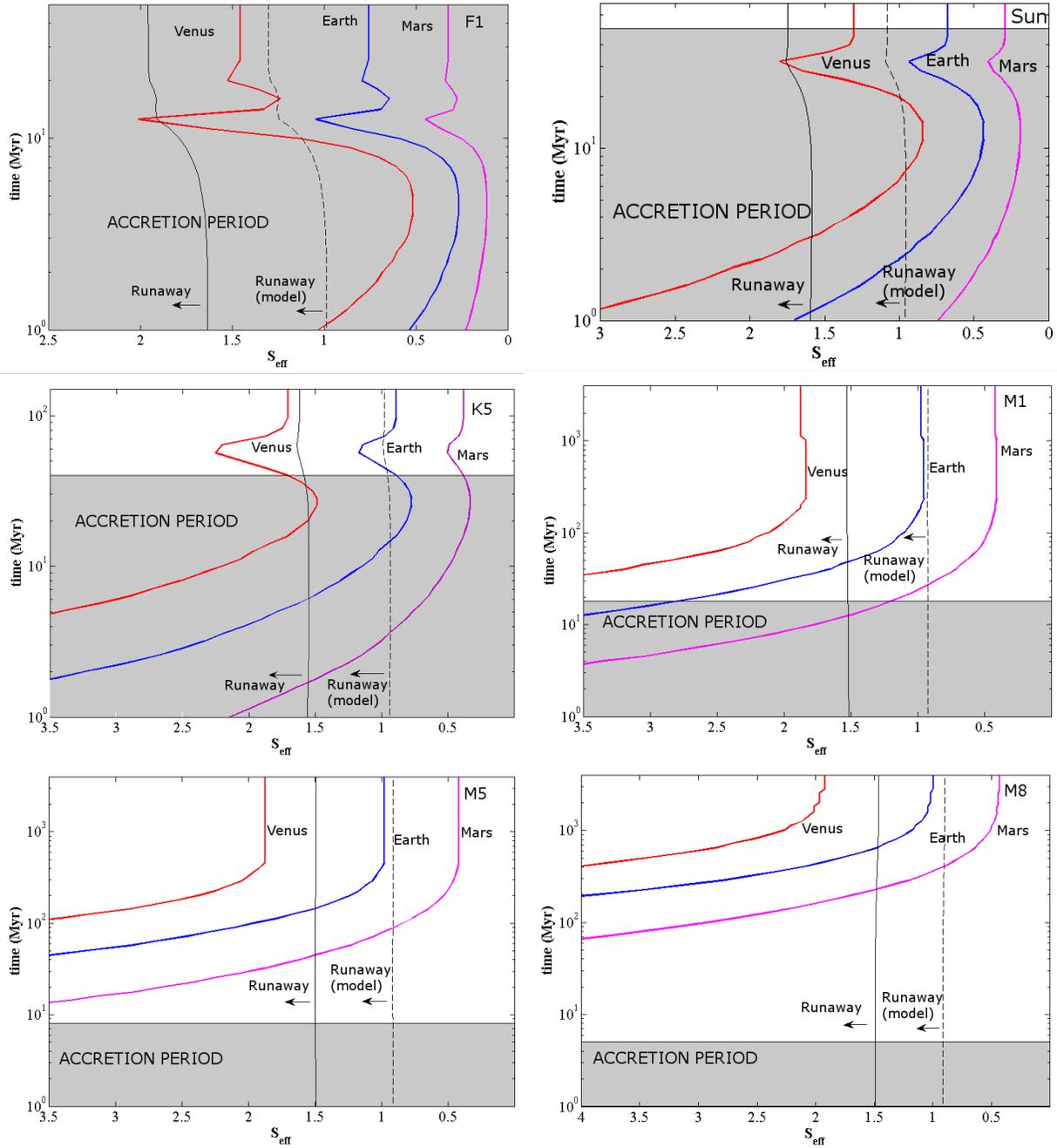

**Fig.4**. Stellar fluxes ($S_{eff}$) received at Venus- (red), Earth- (blue) and Mars-equivalent (purple) distances for F1, solar, K5, M1, M5 and M8-stars. The inner edge is shown for the empirical (solid) and modeled (dashed vertical line) HZ.

*4.2 Lifetime, observability and water retention during the pre-main-sequence*

The short pre-MS lifetime of ~25 – 100 Myr for F- K stars only permits targeted remote observations of young star clusters (e.g. Zuckerman et a., 2004; Malo et al., 2013). Also, the pre-MS lifetimes for M-stars range from about 200 million years (M1) to 2.5 billion years (M8), providing more time to search for potentially habitable environments. Moreover, because pre-MS HZ distances are larger than those for the MS HZ, planets in the pre-MS HZ around cool stars will be easier to resolve with next-generation telescopes (i.e. GMT (10 mas, 1 micron), TMT, and E-ELT(<10 mas, 13 micron)(e.g. Lloyd-Hart et al., 2006; Wright et al., 2010, Kasper et al., 2010). For example, the pre-MS HZ angular resolution for the 40 Myr AP Col (M 4.5, 8.4 pcs away)(Riedel et al., 2011) is ~8 – 24 mas.

Our results (Figs. 3-4) suggest that planets later located in the MS HZ orbiting stars cooler than ~ a K5 may lose most of their initial water endowment. However M-dwarf MS HZ planets could acquire much more water during accretion than did Earth (Hansen, 2014), or water-rich material could be brought in after accretion, through an intense late heavy bombardment (LHB) period similar to that in our own Solar System (Hartmann 2000;Gomes et al.2005). The second mechanism requires that both the gas disk has dissipated and the runaway greenhouse stage has ended. Volatile-rich planets could also migrate or scatter into the pre-MS or MS HZ (e.g. Lissauer et al. 2007). Note that our calculated water losses are upper limits. These would decrease (see eq.1) if, e.g. the stellar extreme UV flux is lower, the initial disk shelters the planet from these fluxes, or $H_2O$ gets entrained from the flow and radiates substantial energy to space, cooling the expansion.

A pre-MS period of up to 195 Myr years for M1 and 2.42 Gyr for M8-stars is available for planets orbiting at distances corresponding to the outer edge of the MS HZ even after such planets would have been initially devolatilized (Fig. 3). Thus, it may be possible that an initially dessicated planet becomes habitable before the star reaches the MS. Moreover, one can speculate that life on a planet could start during the pre-MS phase and move to the subsurface (or underwater) as the star's luminosity decreases.

## 5. CONCLUSION

We calculate the pre-main-sequence HZ for F to M stars. As a result of high initial stellar luminosities, this pre-main-sequence HZ is located farther from the star than is the case for the Main Sequence HZ. The coolest stars can potentially provide habitable conditions for up to 2.5 billion years. Planetary systems around such cool stars can be more easily resolved by upcoming telescopes because of larger planet-star separation than during the MS period.

Accreting planets that are later located in the MS HZ orbiting stars of stellar type K5 and cooler receive stellar fluxes that exceed the runaway greenhouse threshold, and thus may lose a substantial part of the water initially delivered to them. Therefore, M-star planets in the HZ need to initially accrete more water than Earth did or, alternatively, have additional water delivered later to remain habitable. Planets in the pre-main-sequence HZ are interesting targets for remote observations to search for habitable environments and assess mechanisms of water delivery.

**Acknowledgements:**The authors acknowledge support by the Simons Foundation (SCOL

**Table 1: Constants to calculate pre-main-sequence HZ boundaries**

| Recent Venus | 0.08 $M_{sun}$ | 0.1 $M_{sun}$ | 0.2 $M_{sun}$ | 0.3 $M_{sun}$ | [a] 0.5 $M_{sun}$ |
|---|---|---|---|---|---|
| A | 0.2093 | 0.2295 | 0.3248 | 0.368 | 0.5358 |
| B | -0.4283 | -0.4190 | -0.4263 | -0.41 | -0.3517 |
| C | 0.003815 | 0.007261 | 0.002682 | 0.03066 | 0.02723 |
|  |  |  |  |  |  |
| **Leconte model** | 0.08 $M_{sun}$ | 0.1 $M_{sun}$ | 0.2 $M_{sun}$ | 0.3 $M_{sun}$ | [a] 0.5 $M_{sun}$ |
| A | 0.2674 | 0.2932 | 0.4153 | 0.4707 | 0.6868 |
| B | -0.4283 | -0.4182 | -0.4260 | -0.4097 | -0.3511 |
| C | 0.004853 | 0.009112 | 0.03432 | 0.03926 | 0.03473 |
|  |  |  |  |  |  |
| **Early Mars** | 0.08 $M_{sun}$ | 0.1 $M_{sun}$ | 0.2 $M_{sun}$ | 0.3 $M_{sun}$ | [a] 0.5 $M_{sun}$ |
| A | 0.5588 | 0.6119 | 0.8626 | 1.085 | 1.407 |
| B | -0.4322 | -0.4237 | -0.4279 | -0.3963 | -0.3542 |
| C | 0.012000 | 0.021360 | 0.07090 | 0.085030 | 0.07292 |
| **Pre-MS Lifetime** | 2.5Gyr | 1.1Gyr | 400Myr | 350Myr | 200Myr |

[a]Error increases after ~160 Myr due to power law fit.